\documentclass[]{article}
\usepackage{subfigure}
\usepackage{graphicx}

\begin{document}

\title{Temperature dependencies of the energy and time resolution of silicon drift detectors}
\maketitle
\begin{center}
B. K. W\"unschek$^1$, Y. Fujiwara$^2$, T. Hashimoto$^2$, R. S. Hayano$^2$, M. Iio$^3$, S. Ishimoto$^4$, T. Ishiwatari$^1$, M. Sato$^2$, E. Widmann$^1$, J. Zmeskal$^1$
\end{center}

\maketitle

\begin{center}
$^1$Stefan Meyer Institute for Subatomic Physics, Austrian Academy of Sciences, Vienna, Austria\\
$^2$Department of Physics, University of Tokyo, Tokyo, Japan\\
$^3$RIKEN Nishina Center for Accelerator-Based Science, Wako, Japan\\
$^4$KEK, Tsukuba / J-PARC, Tokai, Japan\\
\vspace{1cm}

\textbf{preprint}
%\textbf{Please cite as:} Philosophical Magazine \textbf{90} 4635 - 4645
%DOI: 10.1080/14786435.2010.482178

\vspace{1cm}
\end{center}

\textbf{keywords:} Silicon Drift Detector, energy resolution, time resolution, temperature dependence
\vspace{1cm}

\begin{abstract}

The response of silicon drift detectors (SDDs), which were mounted together with their preamplifiers inside a vacuum chamber, was studied in a temperature range from 100 K to 200 K. In particular, the energy resolution could be stabilized to about 150 eV at 6 keV between 130 K and 200 K, while the time resolution shows a temperature dependence of T$^{3}$ in this temperature range. To keep a variation of the X-ray peak positions within 1 eV, it is necessary to operate the preamplifier within a stability of 1 K around 280 K. A detailed investigation of this temperature  influences on SDDs and preamplifiers is presented. 

\bigskip

\end{abstract}

\section{Introduction} \label{intro}

Silicon drift detectors (SDDs) were initially developed by E. Gatti { \it et al.} \cite{Gatti} as position sensitive detectors. By changing the surface structure \cite{Rehak,Kemmer,Lechner} they also have excellent features in low energy X-ray spectroscopy. This is on the one hand due to their good energy resolution resulting from the small sized anode which provides a small capacitance in spite of a large active area and their good time resolution allowing to operate SDDs at high rates. On the other hand the thickness of the active layer is optimized in order to fully absorb low-energy X-rays ($< 10$ keV). Inside the SDD, the photoelectrons produced by an incoming radiation are collected in the central anode guided by the drift field generated between the center and the edges of the detector. The charges are amplified in a Field Effect Transistor (FET) close to the anode, which also generates the output signal.\\
SDDs have recently been used for X-ray spectroscopy applications such as Proton Induced X-ray Emission (PIXE) \cite{Milota}. Also for kaonic atom X-ray spectroscopy, several types of SDDs with a large active area have been developed and used e.g. in the KEK PS-E570 experiment \cite{Okada} and the SIDDHARTA experiment \cite{Siddharta,Ishiwatari,Bazzi}. They will also be used in the J-PARC E17 experiment, where the 3d$\rightarrow$2p radiative transition in liquid kaonic $^3$He is planned to be measured with an accuracy of 2 eV at $6.2$ keV \cite{Proposal}.\\
In the E17 setup, the SDDs are installed close to the target cell covering a large solid angle and operated at low temperatures in order to reduce thermal radiation to the helium target (kept at 1.4 K). Due to space limitations the preamplifiers have to be installed inside the target chamber, connected via cables to the SDDs. Good energy and time resolutions for all SDDs are required to fulfill the goal of measuring with a precision of 2 eV. \\
Under the described experimental conditions, the response of SDDs is poorly known. Thus, a key point for the success of the E17 experiment was a study in advance of the detectors' response in a setup similar to the E17 setup. In particular, a study of the SDDs' energy and time resolution dependence on temperature was required before installing the devices into the E17 setup. The experimental vacuum chamber, where both SDDs and their associated preamplifiers were installed, is described in section \ref{setup}. The discussion of the temperature dependence of energy and time resolution is then presented in section \ref{test}.\\

\section{Experimental method} \label{exp}

\subsection{Detectors and preamplifiers in the setup} \label{setup}

\begin{figure}[htbp]
  \centering
  \includegraphics[scale=0.7]{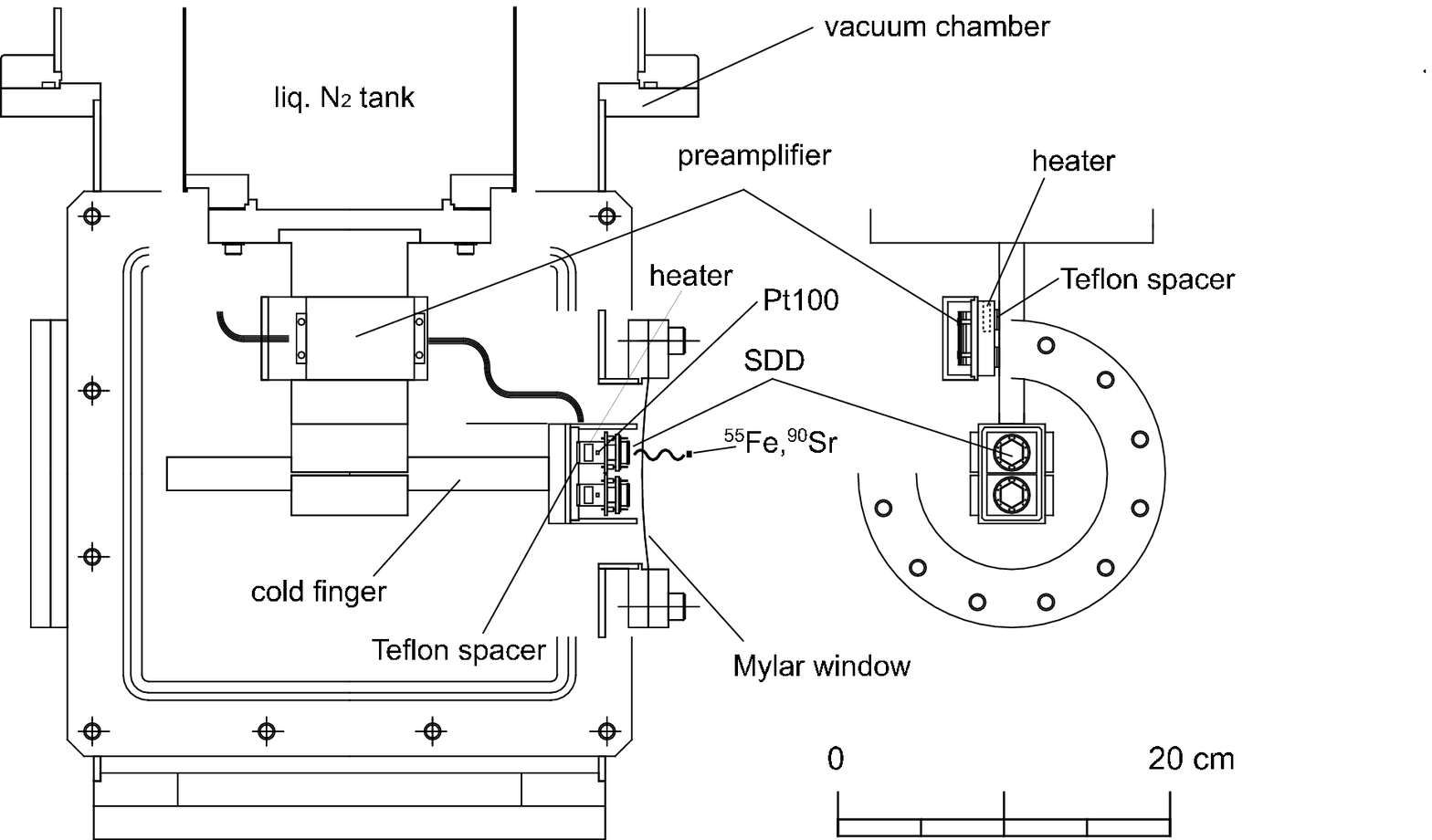}%
  \caption{Schematic picture of the chamber from side view (left) and front view (right). Preamplifiers and detectors inside the chamber were connected conductively to a copper cold finger, cooled by a tank filled with liquid nitrogen. Outside the chamber, radiation sources and - for the determination of the SDD's time resolution - a PIPS detector were fixed in front of the Mylar window, aligned to the SDDs. Between SDDs and sources, the chamber was covered hermetically with a window (mylar, 100 $\mu$m thickness).}
  \label{pic_scheme}
\end{figure}

Altogether nine SDDs\footnote{Type Vitus-SDD "V SOCKEL 100+ KER45", without window and collimator, absorption depth: 450 $\mu$m, active area: 100 mm$^2$} and twelve optical-feedback preamplifiers\footnote{reset type "EPCB-VRPA-S", operating at 12 Volts} made by KETEK were available for all measurements. Since no effective differences between the individual SDDs or preamplifiers were observed, the presented results arising from various combinations can be taken as representative for a typical SDD-preamplifier-pair. Additionally, for the determination of the SDDs' time resolution a PIPS detector\footnote{Passivated Implanted Planar Silicon detector, FD300 series (FD300-17-500rM), fully depleted, thickness: 500 $\mu$m,active area: $300$ mm$^2$} made by Canberra was used.\\ 
In the preparatory stage, a fine tuning of the applied SDD voltages was performed. Optimum voltage values providing best achievable energy resolution, which is stable against voltage fluctuations of several Volts, were kept for all further measurements. \\
The experimental chamber is presented schematically in fig. \ref{pic_scheme}. An aluminum housing containing an SDD and one containing a preamplifier were mounted on a copper cold finger, connected conductively to a tank filled with liquid nitrogen (at 77 K). The preamplifier and the SDD were connected via cables with a length of $\sim 40$ cm. The SDD fixed on aluminum brackets inside its housing was placed very close ($\sim 5$ mm) in front of a 100 $\mu$m thick Mylar window. For energy or time resolution studies, respectively, an X-ray source ($^{55}$Fe) or an electron source ($^{90}$Sr) was placed outside in front of the Mylar window and aligned to the detectors. In particular for the time resolution measurements the PIPS detector was located between window and radiation source. \\
Each detector bracket as well as a chip on each preamplifier were equipped with temperature sensors (PT100). Since detectors are connected conductively to their brackets, the temperature obtained by those sensors can be taken as the temperature on the device. Furthermore, heaters for temperature control were placed on SDD brackets and preamplifiers. \\ 

\subsection{Data taking} \label{daq}

\begin{figure}[htbp]
  \centering
  \includegraphics[scale=0.5]{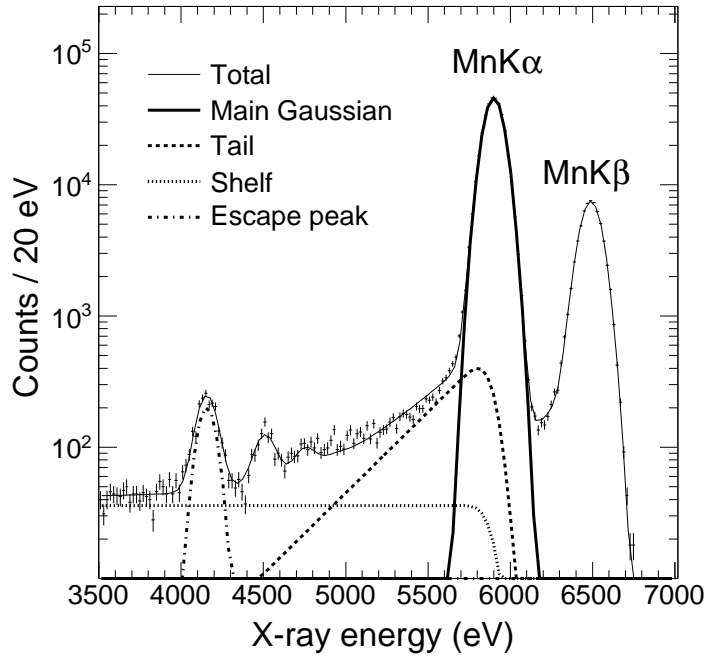}%
  \caption{ADC spectrum of a typical SDD, taken with an $^{55}$Fe source. The peaks, from low to high energy, correspond to: Mn-K$_{\alpha}$ escape peak, Ti-K$_{\alpha}$ peak due to impurities in the surrounding material, Mn-K$_{\beta}$ escape peak, Mn-K$_{\alpha}$ fitted with a Gaussian, a shelf and a tail function, \cite{Campbell}, and finally Mn-{K$_\beta$}. The majority of the detected X-rays contributes to the main Gaussian peak. }
  \label{pic_spectr_er}
\end{figure}

In the X-ray studies, an SDD monitored the radiation from an $^{55}$Fe source. After a first amplification in the preamplifier the output signal was led through a low-pass filter with a cutoff frequency of 3 kHz. The height of the corresponding Gaussian signal, provided by a shaping amplifier\footnote{CAEN N568b}, was measured with an ADC\footnote{TKO Peak Hold ADC, 12 bit, range 2.5V}. By fitting the Mn-K{$_\alpha$} line at $5.9$ keV in the resulting spectra with a Gaussian distribution, a tail function and a shelf function (according to ref. \cite{Campbell}), its peak position and energy resolution (FWHM) were obtained. A typical spectrum with the converted energy on the abscissa is shown in fig. \ref{pic_spectr_er}.

\begin{figure}[htbp]
  \centering
  \includegraphics[scale=0.7]{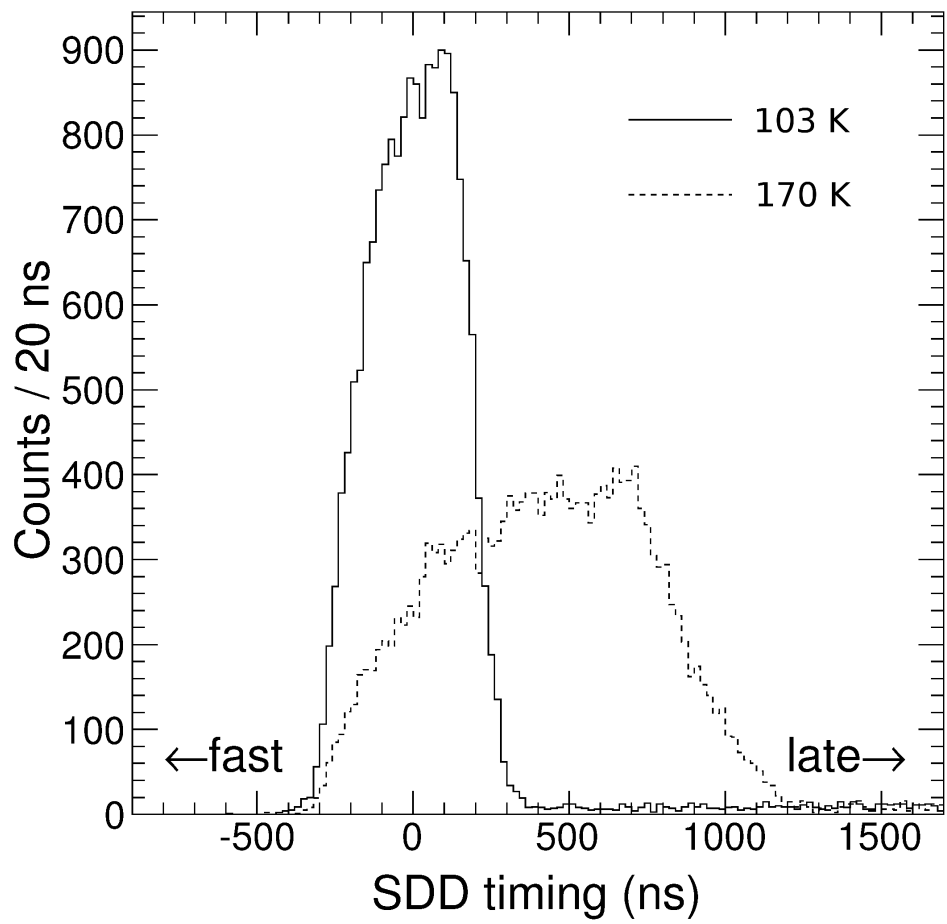}%
  \caption{TDC spectrum of the coincidence signal from PIPS detector and SDD at 103 K (solid line) and 170 K (dashed line), obtained with electrons from $^{90}$Sr.}
  \label{pic_spectr_tr}
\end{figure}

For the electron measurements, a coincidence measurement with a PIPS detector (kept at room temperature) and an SDD (inside the vacuum chamber) was arranged. Due to their construction method, PIPS detectors have even at room temperature an excellent time resolution of $\triangle t_{PIPS} \approx 50$ ns\footnote{The PIPS detector's time resolution was measured with a scintillator in a similar experimental setup.}. The coincidence signal was derived by a TDC\footnote{TKO TDC, 12 bit, range 5 $\mu$s}, a typical time spectrum is presented in fig. \ref{pic_spectr_tr}. The SDD events were recorded in the ADC simultaneously to TDC measurements which allowed to correct a time jitter in electronic devices due to different X-ray energies.

The FWHM of the coincidence signal gives the measured time resolution $\triangle t_{meas}$, which can be expressed as the quadrature sum of all contributions. In our experiment the main contributions arise from the SDD and the PIPS detector: 
	\begin{equation}
	\triangle t_{meas} = \sqrt{{\triangle t_{SDD}}^2+{\triangle t_{PIPS}}^2} \label{equ_trcoll}
	\end{equation}
	with $\triangle t_{PIPS}$ as time resolution of the PIPS detector and $\triangle t_{SDD}$ as time resolution of the SDD, the latter was then derived and used for further calculations. \\
Note that the peak width in fig. \ref{pic_spectr_tr} is a measure for the spreading of the $^{90}$Sr-electron drift times inside the SDD. Their drift paths (and therefore drift times) to the anode are dependent on the incident hit position, because the anode is centrally located on the silicon wafer. Since the electrons uniformly hit the detector surface, the peak width is the sum over all electron drift times.

\section{Temperature studies with SDDs} \label{test}

\subsection{Preamplifier temperature dependence} \label{preamp}

\begin{figure}[htbp] 
  \centering
  \includegraphics[scale=0.35]{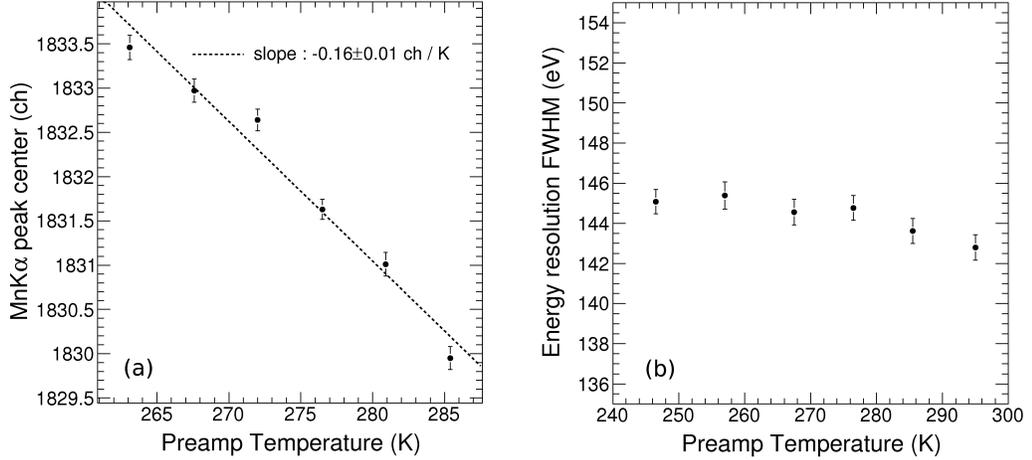}%
  \caption{(a): Peak center shift in units of channels. The linear effect of 0.16 channels per Kelvin corresponds to a shift of $0.5$eV/K. (b): Energy resolution at $5.9$ keV as a function of preamplifier temperature.}
  \label{pic_preampmerge}
\end{figure}

Because the preamplifiers were used in vacuum, the influences of different preamplifier temperatures on the energy resolution and X-ray peak positions were determined.\\
The dependence of peak positions on the preamplifier temperature is shown in fig. \ref{pic_preampmerge}(a), where the center of the Mn-K{$_\alpha$} line is plotted for different temperature values. Statistical error bars are shown. An almost linear negative shift of $0.5$ eV/K within a temperature range from 260 K to 285 K was obtained. Due to this result the preamplifier temperature has necessarily to be controlled more precisely than 1 K in order to keep the peak center stable within 1 eV. Furthermore, below preamplifier temperatures of 270 K, the functionality of the preamplifier is reduced hence they were kept at temperatures above 270 K. Both effects, the reduced functionality as well as the peak shifts are not yet understood completely and need further investigation.\\
The energy resolution in turn, presented as a function of preamplifier temperature in fig. \ref{pic_preampmerge}(b), shows no observable effect. The fluctuations are less than 3 eV within a temperature range of 260 K to 290 K. 

\subsection{SDD temperature dependence} \label{sddtemp}

\begin{figure}[htbp] 
  \centering
  \includegraphics[scale=0.7]{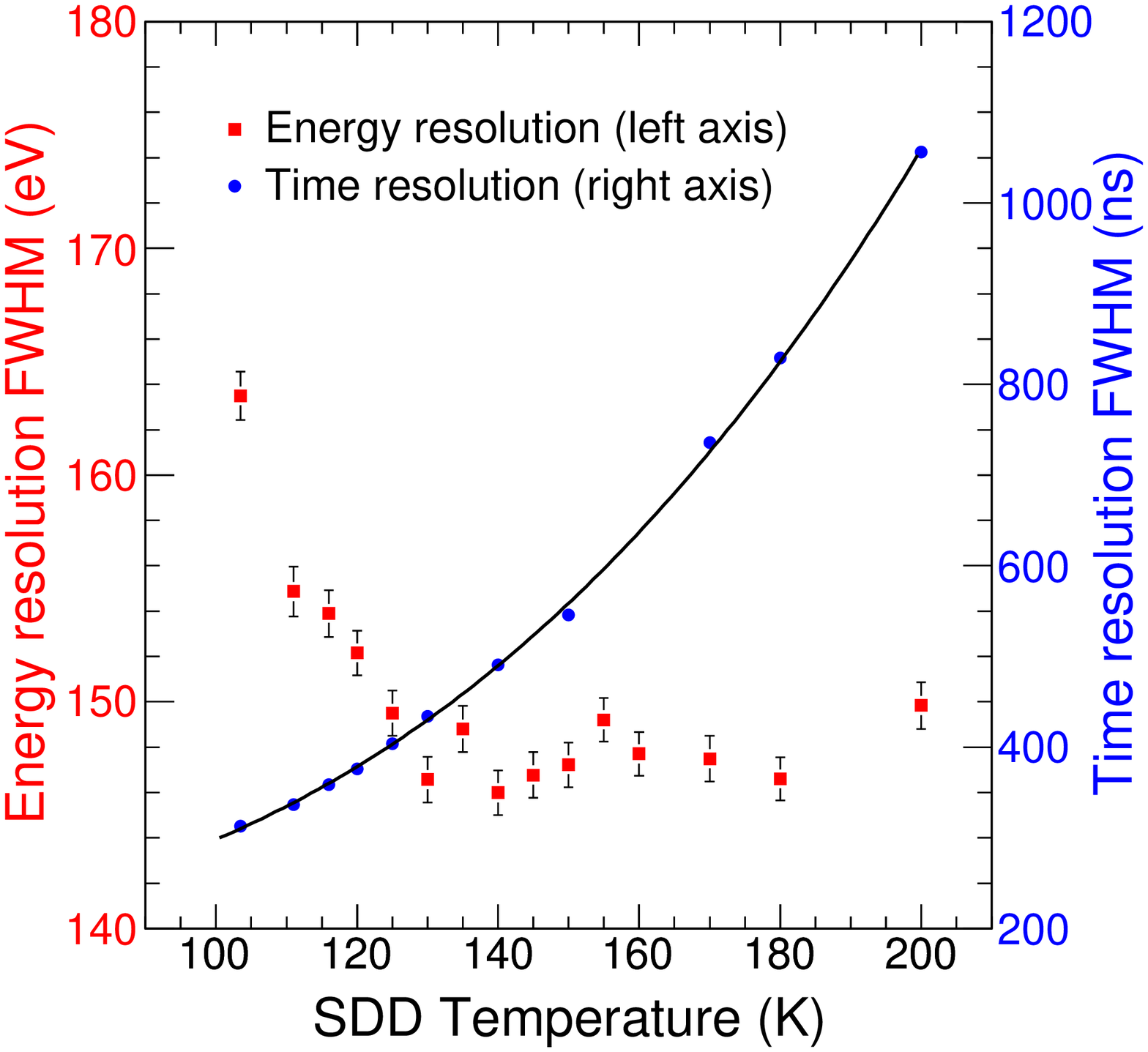}%
  \caption{Energy resolution in eV (squares) and time resolution in ns (triangles) vs. temperature on SDD. The fit corresponds to a $T^n$ function, for details see text.}
  \label{pic_resolutions}
\end{figure}

The energy resolution of SDDs was studied by varying the temperatures from 100 K (which was the lowest value we managed to achieve) to 200 K. The squares and the left axis in fig. \ref{pic_resolutions} correspond to the energy resolution of an SDD in units of eV at $5.9$ keV with statistical errors. The energy resolution is minimized to about 150 eV for SDD temperatures from 130 K to 200 K. Within this temperature range, the energy resolution is stable, but it becomes worse below 120 K. This effect might be attributed to operating limits of the FET. 

\begin{figure}[htbp] 
  \centering
  \includegraphics[scale=0.5]{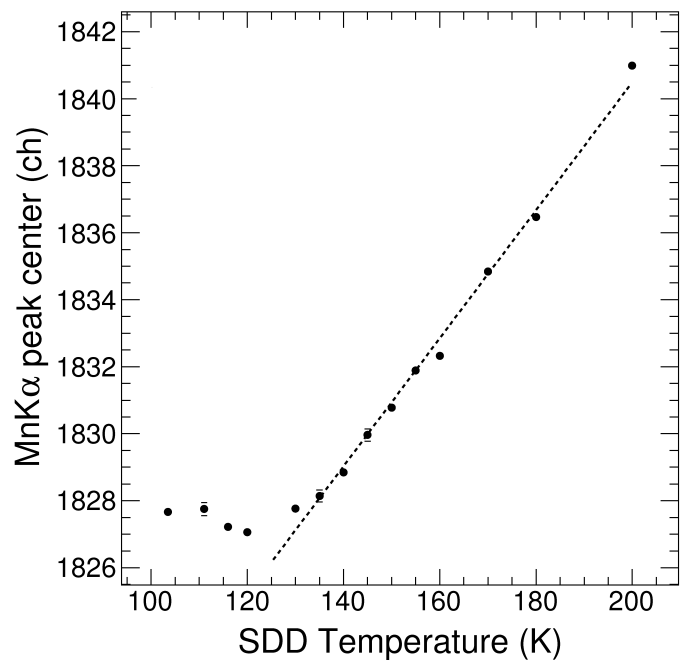}%
  \caption{Mn-K$_{\alpha}$ peak center as a function of temperature on SDD. The linear shift occurring above 130 K and represented by the dashed line is due to the temperature behavior of electron-hole pair creation energy in silicon \cite{Mazziotta}. The shift of $0.19$ channels/K is equivalent to $0.6$ eV/K.}
  \label{pic_sddpeakfit}
\end{figure}

The time resolution of an SDD, $\triangle t_{SDD}$, was measured as a function of its temperature, presented on the right axis (circles) in fig. \ref{pic_resolutions}. It improves with decreasing temperature due to the reduction of thermal noise. We investigated this temperature behavior by comparing $\triangle t_{SDD}$ to the electron mobility in silicon because the time resolution is a measure for the electron travel time in silicon, reciprocally related to the electron drift velocity and the latter in turn dependent on the electron mobility. According to ref. \cite{Seeger}, the electron mobility in silicon is proportional to the temperature by $T^{-n}$, with n around 2.5 for temperatures between 100 K and 300 K. Therefore, our values for $\triangle t_{SDD}$ were fitted with a $T^n$ function (solid line in fig. \ref{pic_resolutions}): 
\begin{equation}
\triangle t_{SDD}(T) = a \cdot \left( \frac{T}{T_0}\right)^{n}+b
\label{equ_muwir}
\end{equation}
with $T_{0} = 300$ K. The fit gives $a=(2942 \pm 63)$ ns, $b=(190 \pm 6)$ ns and, consistently with reference \cite{Seeger}, $n=(3.00 \pm 0.06)$. In order to understand the fit values $a$ and $b$, a brief estimate of the relation of electron mobility and time resolution in our detectors follows: \\
The electron drift velocity $v$ and the electron mobility $\mu$ are related via the drift field $E$: 
	\begin{equation}
	v = E \cdot \mu
	\end{equation}
For the sake of simplicity, the drift field $E$ in a two-dimensional and circular-shaped assumed SDD with radius $R$ can be described by the ratio of the voltage difference between center and edge to the maximum electron drift path $R$ (i.e. radius). Furthermore, the electron drift velocity can also be expressed as the ratio of its drift path in the SDD to its drift time. In this simple model, the difference between the maximum and the minimum electron drift time is taken as a measure for the time resolution $\triangle t_{SDD}$. From this it follows that the { \it relation of time resolution and electron mobility} can be expressed via: 
	\begin{equation}
	\triangle t_{SDD}(T) = \frac{R}{\mu_{0} \cdot E} \cdot {\left( \frac{T}{T_{0}} \right)}^{n} + c 
	\label{equ_mut}
	\end{equation}
with $\mu_{0}$ as the electron mobility at the reference value $T_{0}$ and c as a constant, representing $\triangle t_{SDD}$ at $T = 0$ K. In comparison with equation \ref{equ_muwir}, it shows that the constant $a$ thus includes geometrical characteristics of the detector, of the electrical field and of material properties. \\
We furthermore compared our results with published values of $\mu$ in silicon ($\mu \approx 1400$ cm$^2$s$^{-1}$V$^{-1}$ at 300 K). Therefore, the time resolution of our SDD at 300 K, according to the fit function in equation \ref{equ_muwir}, was calculated. In our SDDs, the maximum drift path is $0.62$ cm and the difference of the applied voltages is around $110$ V. This gives in accordance with equation \ref{equ_mut} an electron mobility for the SDD of about $1100$ cm$^2$s$^{-1}$V$^{-1}$ at 300 K.\\
\\
Also the Mn-K$_{\alpha}$ peak center was plotted as a function of SDD temperature in fig. \ref{pic_sddpeakfit}. Below 130 K, no unambiguous tendency is observable, which again might be caused by a reduced performance of the FET at low temperatures. Above 130 K, the peak centers show an increasing behavior with increasing temperature. The linear fit in fig. \ref{pic_sddpeakfit} delivers a gradient of $0.19$ channels per Kelvin, which corresponds to $0.6$ eV/K. \\
To understand this gradient, the behavior of the peak shifts was compared to the temperature dependence of the electron-hole pair creation energy in silicon. The SDDs' output signal height for an X-ray, from which the ADC derives a corresponding peak center, is proportional to the energy of this incident X-ray and therefore to the number of the created electrons. The number of created electrons on the contrary is inversely proportional to the electron-hole pair creation energy in silicon. According to Mazziotta et al. \cite{Mazziotta}, who studied the temperature dependence of this creation energy with Monte Carlo simulations, it decreases with $10^{-4}$ eV per Kelvin. Thus, the X-ray peak centers also might be shifted with increasing temperature by a factor of $10^{-4}$ of the incident X-ray which is consistent with the measured peak shift of $0.6$ eV/K corresponding to $10^{-4}$ of the X-ray energy ($5.9$ keV) per Kelvin. 

\section{Conclusions} \label{conclusio}

We investigated the energy resolution and the time resolution of an SDD, as well as the peak center of the Mn-K$_{\alpha}$ line at $5.9$ keV as a function of preamplifier and SDD temperatures. Preamplifier temperature changes do not have a significant influence on energy resolution, but a fortiori an effect on peak positions. In order to avoid peak position shifts of more than 0.5 eV and a reduced functionality of the preamplifier, they were stabilized with a precision better than 1 K above 270 K. A temperature between 270 K and 280 K was chosen for all preamplifiers.\\
The time and energy resolutions and also the peak center positions were determined as a function of SDD temperature. The time resolution was measured via a coincidence measurement with a PIPS detector. It improves with decreasing temperature, which is an effect of reduced noise at low temperatures. The tendency of the time resolution temperature dependence can be understood from the temperature dependence of the electron mobility in silicon. Our results show a $T^n$ (with $n=3.00$) tendency in temperature which fits to theoretically calculated values. Furthermore, it was found that the energy resolution is constant with approximately 150 eV at $5.9$ keV from 130 K to 200 K. The behavior of the peak centers with respect to SDD temperature shows a linear upwards shift above 120 K, induced by the temperature dependent electron-hole pair creation energy in silicon. In both studies, peak center stability and energy resolution as a function of SDD temperature, a different behavior of the SDDs below 120 K could be observed, which might be attributed to operating limits of the FET. In particular, energy resolution becomes slightly worse, but still tolerable below 120 K. \\

{\bf Acknowledgments}\\
This work was supported by the Austrian Science Fund (FWF) [P20651-N20] and the Grant-in-Aid 20002003 for Specially Promoted Research, MEXT, Japan. We also acknowledge the technical support of SMI.

\bibliographystyle{tphl}

\begin{thebibliography}{00}
\bibitem{Gatti} E. Gatti and P. Rehak, Nucl. Instrum. Meth. A. 225 (1984) 608-614
\bibitem{Rehak} P. Rehak, {\it et al }., Nucl. Instrum. Meth. A. 235 (1985) 224-234
\bibitem{Kemmer} J. Kemmer, {\it et al }., Nucl. Instrum. Meth. A. 253 (1987) 378-381
\bibitem{Lechner} P. Lechner, {\it et al }., Nucl. Instrum. Meth. A. 458 (2001), 281-287
\bibitem{Milota} P. Milota, {\it et al }., Nucl. Instrum. Meth. B. 266 (2008) 2279-2285
\bibitem{Okada} S. Okada, {\it et al }., Phys. Lett. B. 653 (2007) 387-391
\bibitem{Siddharta} M. Bazzi, {\it et al }., Phys. Lett. B. 681 (2009) 310-314
\bibitem{Ishiwatari} T. Ishiwatari, Nucl. Instrum. Meth. A. 581 (2007) 326-329
\bibitem{Bazzi} M. Bazzi, {\it et al }., Nucl. Instrum. Meth. A in press (2010) doi:10.1016/j.nima.2010.06.332
\bibitem{Proposal} R.S. Hayano, Nucl. Phys. A. 827 (2009), 324c-326c
\bibitem{Campbell} J.L. Campbell, {\it et al }., X-ray Spectrometry 30 (2001) 230-241
\bibitem{Seeger} K. Seeger, Semiconductor Physics, Springer, 1973
\bibitem{Mazziotta} M.N. Mazziotta, Nucl. Instrum. Meth. A. 584 (2008) 436-439
% \bibitem{}

\end{thebibliography}

%%%%%%%%%%%%%%%%%%%%%%%%%%%%%%%%%%%%

\label{lastpage}

\end{document}